\documentclass[11pt,twoside]{article}
\usepackage{asp2014}

\aspSuppressVolSlug
\resetcounters

\begin{document}

\title{SciCodes: Astronomy Research Software and Beyond}

\author{Alice~Allen$^1$ }
\affil{$^1$Astrophysics Source Code Library/University of Maryland, College Park, MD, USA; \email{aallen@ascl.net}}

\paperauthor{Alice~Allen}{aallen@ascl.net}{0000-0003-3477-2845}{Astrophysics Source Code Library/University of Maryland, College Park}{Astronomy Department}{College Park}{}{MD}{USA}




  
\begin{abstract}

The Astrophysics Source Code Library (ASCL ascl.net), started in 1999, is a free open registry of software used in refereed astronomy research. Over the past few years, it has spearheaded an effort to form a consortium of scientific software registries and repositories. In 2019 and 2020, ASCL contacted editors and maintainers of discipline and institutional software registries and repositories in math, biology, neuroscience, geophysics, remote sensing, and other fields to develop a list of best practices for these research software resources. At the completion of that project, performed as a Task Force for a FORCE11 working group, members decided to form SciCodes as an ongoing consortium. This presentation covered the consortium's work so far, what it is currently working on, what it hopes to achieve for making scientific research software more discoverable across disciplines, and how the consortium can benefit astronomers.
  
\end{abstract}

\section{Introduction}
The Astrophysics Source Code Library (ASCL ascl.net) is a free online registry for source codes of interest to astronomers and astrophysicists, including solar system astronomers. It registers scientist-written software used in research that has appeared in, or been submitted to, peer-reviewed publications; it can also serve as a repository. The ASCL assigns a unique identifier to its entries, and improves research transparency and reproducibility by making these codes more discoverable for examination.

A broad look at research software shows that codes developed in one discipline may also be useful in another. For example, WND-CHARM \citep{Shamir2008}, first used in X-ray imaging analysis \citep{Shamir2009}, has also been used in galaxy morphology research \citep{Kuminski2014}. emcee \citep{Foreman-Mackey2013}, first written for use in astrophysics and now widely used in the field, has also been used in other disciplines, including materials science \citep{Paulson2019}, Ebola virus research \citep{Liao2020}, and ecology \citep{Mairet2021}. 

It is not surprising that other disciplines also have registries or repositories for software used in those fields, and that these registries have many of the same goals, concerns, and interests as the ASCL. Among these are software citation and ensuring code authors receive credit for the computational methods that enable research, curation of the records they hold, adopting and adapting practices that support and improve stewardship of the resources themselves, and increased software discoverability both for possible (re)use and to support research by making these methods available. 

That these various registries participate in organizations such as the Research Data Alliance \citep{Berman2020} and FORCE11 \citep{bourne2012} demonstrates that they share a number of interests and goals. In late 2018, the ASCL, as a member of the FORCE11 Software Citation Implementation Working Group,\footnote{https://github.com/force11/force11-sciwg} suggested the group form a task force to develop a list of best practices for research software registries. This provided an opportunity for managers and editors of these resources to meet regularly and work toward a common goal. These meetings, a Sloan-funded workshop in November 2019,\footnote{https://asclnet.github.io/SWRegistryWorkshop/} and subsequent work led to the release of best practices for such resources in late 2020 \citep{BestPractices2020}. At the conclusion of that project, the task force's work done, the participants decided to keep meeting and formed the SciCodes consortium.

\section{SciCodes}
SciCodes\footnote{https://scicodes.net} is a consortium of scientific software registries and repositories. The consortium meets monthly, offering two meetings on the same day to enable broad geographic participation. The consortium's first meeting was in February 2021. In this formative first year, the group was led by three co-chairs, Michael Hucka and Thomas Morrell, both at Caltech, and this author. Participating resources, listed in the Table\ref{participants} in the Appendix, were polled on various governance issues, and the group chose, for subsequent years, to have two co-chairs with staggered terms, elected in October and with their terms to start the following January. The group also developed and ranked a list of goals to guide the consortium's work.

Meetings typically offer a short presentation, which is recorded and made available online,\footnote{https://scicodes.net/presentations/} and discussion and progress reports on high-priority goals. These goals include: 
\begin{itemize}
    \item Enable searching across multiple software registries
    \item Speed adoption of CodeMeta \citep{Jones2017} and CFF \citep{Druskat2019} standards to improve software citation and discoverability
    \item Strengthen our resources through implementation of identified best practices
    \item Keep up with and share advances and ideas
    \item Discuss challenges and solutions to common issues that arise in managing our resources
\end{itemize}

\section{How SciCodes benefits astronomers}
The SciCodes consortium's work will offer numerous benefits to astronomy. By making scientific software more discoverable across disciplines, SciCodes will enable astronomers in finding useful research software regardless of which discipline it was developed in; further, it will help make astronomy software more discoverable to those in other fields. This can improve the efficiency of research, and increase citation of computational methods. SciCodes also works to strengthen each of the participant registries and repositories, making the workings of these resources more transparent to software authors, indexers, and journal data editors, and providing information that reflects on their use of best practices. 

\section{Acknowledgements}
The ASCL thanks the Alfred P. Sloan Foundation for its generous support of the Scientific Software Registry Collaboration Workshop, and Michigan Technological University, the University of Maryland (College Park), and the Heidelberg Institute for Theoretical Studies for their generous ongoing support. 

\bibliography{X9-012}
\newpage
\appendix
\section{Appendix}
\begin{table}[!h]
\caption{SciCodes Participating Registries and Repositories}
\label{participants}
\smallskip
\begin{center}
{\small
\begingroup
\renewcommand{\arraystretch}{1.2}
\begin{tabular}{ll}

\hline
\textbf{Resource}  & \textbf{URL} \\ 
\hline

Astrophysics Source Code Library\\(ASCL)   & \url{https://ascl.net} \\ \hline
Australian Research Data Commons \\(ARDC)   &  \url{https://ardc.edu.au/} \\ \hline
Biological General Repository for\\Interaction Datasets \\(BioGRID)   & \url{https://thebiogrid.org/} \\ \hline
CaltechDATA  & \url{https://data.caltech.edu/} \\ \hline
Computational Infrastructure for \\ Geodynamics\\(CIG)  & \url{https://geodynamics.org/} \\ \hline
DOECODE  & \url{https://www.osti.gov/doecode/} \\ \hline
ELIXIR bio.tools   & \url{https://bio.tools/} \\  \hline
HAL  & \url{https://hal.archives-ouvertes.fr/} \\ \hline
Harvard Dataverse  & \url{https://dataverse.harvard.edu/} \\ \hline
Model Integration \\(MINT)   & \url{http://mint-project.info/} \\ \hline
ModelDB   &  \url{https://senselab.med.yale.edu/modeldb/} \\ \hline
Network for Computational Modeling \\ in Social and Ecological Sciences\\(CoMSES)  & \url{https://www.comses.net/} \\ \hline
Oak Ridge National Laboratory \\ Distributed Active Archive \\ Center for Biogeochemical  Dynamics \\ (ORNL DAAC) & \url{https://daac.ornl.gov/} \\ \hline
Ontosoft  & \url{http://www.ontosoft.org/} \\ \hline
PhysioNet  & \url{https://physionet.org/} \\ \hline
Remote Sensing Code Library\\(RSCL)  & \url{https://rscl-grss.org/} \\ \hline
SBGrid   & \url{https://sbgrid.org/} \\ \hline
SciCrunch  & \url{https://scicrunch.org/} \\ \hline
simTK  & \url{https://simtk.org/} \\ \hline
Software Heritage  & \url{https://www.softwareheritage.org/} \\ \hline
swMATH  & \url{https://swmath.org/} \\ \hline
Systems Biology Markup Language\\(SBML)  & \url{http://sbml.org/Main\_Page} \\ \hline
Zenodo   & \url{https://zenodo.org/} \\ \hline

\end{tabular}
\endgroup
}

\end{center}
\label{undefined}
\end{table}
\normalsize


\end{document}